\documentclass[
 reprint,
 amsmath,amssymb,
 aps
]{revtex4-2}
\usepackage{xcolor}
\usepackage{graphicx}
\usepackage{dcolumn}
\usepackage{bm}

\usepackage{amsmath}
\usepackage[colorlinks=true, allcolors=blue]{hyperref}

\begin{document}
\title{Theoretical description of proton--deuteron interactions using exact two-body dynamic of femtoscopic correlation method}
\author{Wioleta Rzesa$^{1}$, Maria Stefaniak$^{2}$, Scott Pratt$^{3}$}

\affiliation{$^{1}$ Warsaw University of Technology, Faculty of Physics, ul. Koszykowa 75, 00-662, Warsaw, Poland\\
 $^{2}$ The Ohio State University, Department of Physics, 191 Woodruff Avenue, Columbus, OH 43210\\
 $^{3}$ Michigan State University, Facility for Rare Isotope Beams, 640 South Shaw Lane, East Lansing, MI 48824 }

\date{\today}


\begin{abstract}
Modeling proton-deuteron interactions is particularly challenging. Due the deuteron's large size, the interaction can extend over several femtometers. The degree to which it can be modeled as a two-body problem might also be questioned.
One way to study these interactions is through femtoscopic correlation measurements of particle pairs, extracting information using available theoretical models. In this work, we examine two approaches for describing proton--deuteron correlations: the Lednick\'{y}-Lyuboshits formalism and full numerical solutions of the Schr\"odinger equation. Our results show that the differences between these methods are significant. Furthermore, we demonstrate that incorporating higher-order partial waves—particularly the p-wave—is essential for accurately capturing the dynamics of proton--deuteron interactions and the full potential of the strong force.
\end{abstract}

\maketitle
\section{Introduction}
One of the most compelling topics in interaction studies in recent years is the investigation of the proton-deuteron (\textit{p-d}) system. This system is of particular interest because it involves interactions between three nucleons as the deuteron is composed of both a proton and a neutron. Therefore, an investigation of the \textit{p-d} system serves as a valuable laboratory for exploring many-body physics. In particular, for this purpose the low-energy scattering wave function can provide invaluable insights. The interactions in \textit{p-d} system has primarily been parameterized based on scattering experiments ~\cite{Oers1967, Arvieux1974, Huttel1983, Kievsky1997, Black1999}, which offer important information on scattering cross sections and phase shifts of interactions. For scattering energies below the deuteron's binding energy of $2.2$~MeV, the phase shifts encapsulate all accessible information from the scattering data.

%


Here, we focus on correlation studies of particles in expanding matter produced in particle and ion collisions, commonly known as the femtoscopy technique ~\cite{femto2decades}, which can be used to extend the analysis of the interactions in the \textit{p-d} system. Initially developed to investigate source properties, femtoscopy has evolved into a powerful tool for detailed experimental studies of interactions at low relative particle momenta ~\cite{kp_in_pp, kp_in_PbPb, barionantibarioninter, STAR:2014shf, 2023138145}. 
In principle, such correlations are sensitive to the scattering wave-function itself.
By measuring two-particle femtoscopic correlations, researchers gain unique insights into the dynamics of particles, including those that are theoretically challenging and not easily accessible through scattering experiments.


Furthermore, femtoscopic correlations have been employed by the authors of Refs~\cite{Mrowczynski:2019yrr, ant:2, ant:3, ant:4} to study the mechanism of cluster formation in heavy-ion collisions, a phenomenon that remains incompletely understood~\cite{prod1, prod2, prod3, prod4}. This is because the femtoscopic source size of particles which are produced in thermal processes~\cite{SHMmodel} differs from that in coalescence~\cite{CoalInHIC} nature. However, to distinguish experimentally between these two scenarios and to extract meaningful information about the strong interaction, it is crucial to have a robust theoretical understanding of how details of the wave function might affect the correlation function.
In this paper, we emphasize the necessity for a careful and precise treatment of calculations involving the wave functions that describe the interactions between protons and deuterons. 
Due to the large size of the deuteron, the \textit{p-d} effective interaction extends several fm, which should make \textit{p-d} femtoscopy especially sensitive to details of the treatment.

\section{Theoretical description of proton--deuteron correlations}

Femtoscopic studies of correlation functions have a long and well-established history, originating from early two-photon correlation studies in astronomy~\cite{hbt1954}. This methodology was subsequently adapted to particle physics, particularly in the study of momentum correlations for small and spatially close particle sources~\cite{Goldhaber1960, KOPYLOV1972, KOPYLOV1974}. When applied to such small radiating sources, femtoscopy is used to explore the space-time evolution of particle emission by measuring the yield of particle pairs as a function of their relative momentum~\cite{sinyukov1, femto2decades}. The theoretical foundation for describing correlations between two particles is commonly expressed through the Koonin-Pratt formula~\cite{Koonin:1977fh, pratt:1986}:
 \begin{equation}
     C(\mathbf{k}^*) = \int S(\mathbf{r^*}) |\Psi\left(\mathbf{r^*},\mathbf{k^*}\right)| \text{d}^3 r^* \,, \label{Eq. Koonin-Pratt}
 \end{equation}
where $*$ denotes variables in the pair-rest frame, $k^* = |\mathbf{p}^*_1-\mathbf{p}^*_2|/2$, and $\mathbf{p}^*_{1,2}$ are the particle momenta, $r^*$ describes the relative distance between the two particles. 

The function $S(\mathbf{r}^*)$ represents the probability that two particles emitted with the same velocity would be separated by a relative distance $r^*$, in the limit that the interaction between the two particles is neglected. The formalism is contingent on the assumption that other sources of correlation, such as energy or momentum conservation are ignored. For heavy-ion sources, or even for higher-multiplicity proton-proton collisions, this treatment should be sound.
The modeled correlation function typically assumes a Gaussian profile for the particle-emitting source ~\cite{Broniowski:2008vp,Acharya:2017qtq,Acharya:2020nyr}. In one-dimensional expression, this takes the form:
\begin{equation}
S(\mathbf{r^*}) \sim \exp\left(-\frac{r^{*2}}{4R^2}\right),
\end{equation}
where $R$ refers to the femtoscopic source size of the analyzed particle pair.

The function $\Psi(\mathbf{k}^*, \mathbf{r}^*)$ represents the wave function, describing the interactions between the two particles. This wave function can either be derived by solving the Schr\"odinger equation for a given potential or obtained through parameterization of the interaction. The determination of the $\Psi(\mathbf{k}^*, \mathbf{r}^*)$ for \textit{p-d} interaction is a complex problem, as the deuteron is a composite particle consisting of two nucleons. Consequently, the interacting system is relatively large, with the deuteron's radius already being approximately $\sim$2.1~fm. The dynamics of this interaction are still debated, as the system can be modeled either as an interaction between a proton and a deuteron (two-body approach) or as an interaction between a proton and the individual nucleons within the deuteron, treated separately (three-body approach).

The nature of \textit{p-d} dynamics has been the focus of recent theoretical~\cite{pd_theory_3body_Viviani2023} and experimental~\cite{pd_ppcol_ALICE} studies, where the \textit{p-d} correlation function was compared using both two-body and three-body approaches. Notably, only the three-body approach successfully reproduced the experimental correlation function. However, this study primarily addressed pp collisions, where because of small femtoscopic sizes ($\sim$1.5~fm) the external proton is basically in the vicinity of the deuteron itself. At such small distances, three-body effects become prominent in the system's dynamics. The three-body effects are expected to diminish in larger systems, such as those created in heavy-ion collisions, where the femtoscopic source sizes are significantly larger (3-7~fm).

A commonly used two-body approach which was already utilised with success to describe the correlation function of several particle pairs, e.g. kaon--proton ~\cite{kp_in_PbPb} or kaon--deuteron~\cite{pd_ppcol_ALICE} is based on the Lednick\'{y}-Lyuboshits (L-L) formalism~\cite{Lednicky1981, nonidentical_femtoscopy_Lednicky:2009zza}. This approach relies on several approximations, enabling the parameterization of the wave function for both Coulomb and strong interactions as:
\begin{equation}
\small
\Psi^{(+)}_{-{\vec{k}^{*}}}({\vec{{r}^{*}}}) = \sqrt{\frac{A_{\rm C} (\varepsilon)}{2}} \left [ {\rm e}^{-{\rm i} \vec{ k}^{*}\cdot{\vec{r}^{*}}} {\rm F}(-{\rm i} \varepsilon, 1,
  {\rm i} \zeta^{+}) + f_{\rm C}(\vec{k}^*)\frac{\tilde{G}(\rho,\varepsilon)}{{r}^*} \right ],
\label{eq:fullpsi2}
\end{equation}
\vspace{0.01cm}\\
\normalsize
where $A_{\rm C}$ is the Gamow factor,
$\varepsilon = 1/(k^{*} a_{\rm C})$, 
$\zeta^{\pm} = k^{*} r^{*} (1 \pm \cos{\theta^{*}})$, 
$\rm F$ is the confluent hypergeometric function, and $\tilde{G}$ is the combination of the regular and singular s-wave Coulomb functions.
The angle $\theta^{*}$  delineates the relationship between the pair's relative momentum and relative position within the pair's rest frame, whereas $a_{\rm C}$ represents the Bohr radius of the pair. The term $f_{\rm C}(k^*)$ is  the scattering amplitude resultant from strong interaction, adjusted to accommodate the Coulomb component:
\begin{equation}
\label{eq:scattering_amplitude}
    f^{-1}_C(k^*) = \frac{1}{f_0} + \frac{1}{2}r_0{k^*}^2-\frac{2}{a_C}h(k^* a_C) -\text{i}k^* a_C,
\end{equation}
where $h(\varepsilon)=\varepsilon^2\sum\limits_{n=1}^{ \infty}[n(n^2+\varepsilon^2)]^{-1}
-\mathrm{{\gamma}} -\ln|\varepsilon|$
($\mathrm{{\gamma}} = 0.5772$ represents the Euler constant), $f_{0}$ is the scattering length of the strong interaction, $r_0$ is the zero effective-range of the interaction.

\begin{figure*}[tbh!]
\centering
\includegraphics[width=0.82\linewidth]{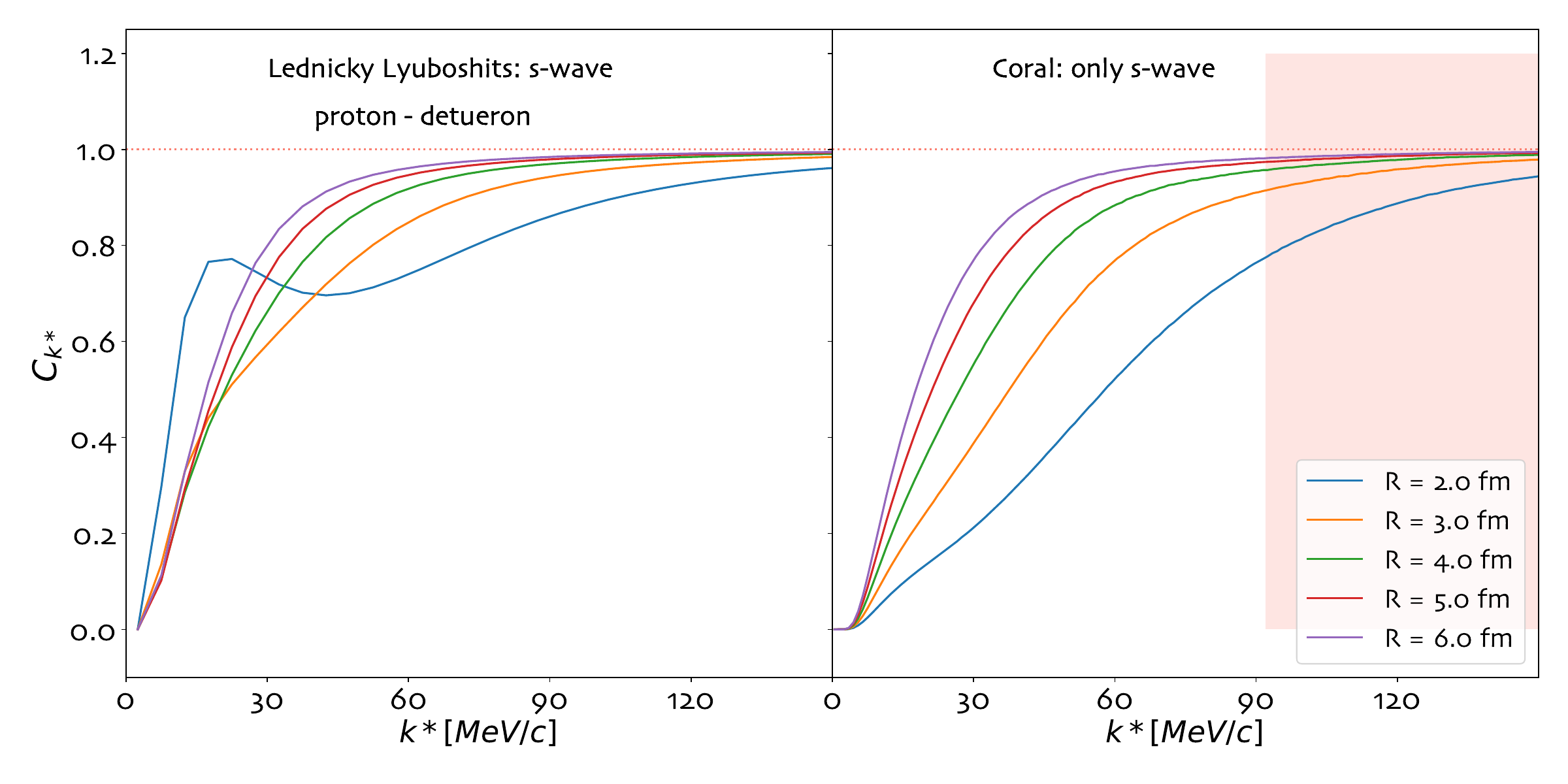}
\caption{The \textit{p-d} correlation functions for various source sizes calculated using: left panel- L-L formalism, right panel - CorAL framework and Schr\"odinger based solution of the wave function. Both approaches account for Coulomb and strong (only s-wave) interactions.}
\label{fig:swave_2panel}
\end{figure*}

The \textit{p-d} interaction is characterised by its description in two spin channels: the doublet with $S$=1/2 and the quartet with $S$=3/2. The two spin states can be accounted by adding the Clebsch-Gordan coefficients~\cite{clebish_gordon} that leads to the sum expressed as:
\begin{equation}
    \label{eq:pdweight}
    C^{total}_{pd} = \frac{1}{3} C^{^2S_{1/2}}_{pd}+\frac{2}{3} C^{^4S_{3/2}}_{pd},
\end{equation}
where $C^{^2S_{1/2}}$ and $C^{^4S_{3/2}}$ represent the correlation function of the two \textit{p-d} spin channels. An example of s-wave correlation functions with scattering lengths~\cite{Black1999} and zero effective-range approximation ($r_{0} $ = 0) calculated using the L-L formalism is shown in left panel of Fig.~\ref{fig:swave_2panel}. The functions characterize a peak structure at $k^* \simeq $20~MeV/$c$ for small femtoscopic sources, which diminishes with increasing system size.

The two-body \textit{p-d} calculations used for comparison with a three-body approach in~\cite{pd_ppcol_ALICE} were also based on the L-L formalism. However, this approach failed to reproduce the \textit{p-d} correlation functions in small femtoscopic sources. This failure is evident because the characteristic large peak at very small $k^*$ $\sim$ 20~MeV/$c$ is not observed in any experimental data ~\cite{pd_ppcol_ALICE, Stefaniak_pd}. Due to the extremely small relative distances between the two particles in pp collisions, the particular behaviour of the L-L approach and lack of agreement with data describing the small femtoscopic sources has been attributed mainly to missing three-body effects, such as Pauli blocking between the two protons. In this work, we also highlight another important limitation of the L-L formalism which could contribute to discrepancies observed in the correlation function. 

The wave function in terms of interaction is a superposition of an incoming plane wave and an outgoing spherical wave. 
The outgoing wave function is disturbed by the interaction, and if Coulomb is neglected, it can be parameterised as:
\begin{equation}\label{eq::LL}
\phi_{k^*}(r) = e^{i\delta}\sin(k^*r+\delta)
\end{equation}
where $r$ is the range of interaction, and $\delta$ is a parameter related to the phase shift caused by the interaction. This approximation is exact only when $r$ is larger than the range of the nuclear reaction $r_0$. For $r<r_0$, the potential no longer vanishes, and the approximation significantly deviates from the true wave function. 
It is important to note that the proposed asymptotic form does not satisfy the boundary condition $\phi(k^*, r) \to 0 \text{ as } r \to 0$, where the true wave function vanishes completely. 
However, if the characteristic source size R is much larger than $r_0$, the expression of Eq.~\ref{eq::LL} aligns well with the true form of $\phi(k^*, r)$ except for a small fraction of the overall source, $r_0^3/R^3$.

Most femtoscopic correlation measurements involving particles such as pions, kaons, or protons correspond to femtoscopic source sizes larger than $r_0$ which for small particles is about 1 fm and therefore such systems can be adequately described using this asymptotic form. However, in interactions involving deuterons, interaction can extend beyond the size of the deuteron. One might say that the proton and deuteron, if they are not squeezed within each others, are separated by $\sim$3 fm, so the range of interaction, $r_{\rm 0}$, might extend beyond that value.
As a result, any interaction between protons and deuterons at distances $r < r_0$ will reveal significant differences between the simplified asymptotic form and the true wave function.

An alternative approach which can be used to describe the correlations involves numerically solving the Schr\"odinger equation utilizing potentials derived from fits to experimentally measured phase shifts~\cite{Black1999, improved_phaseshift}. Using these phase shifts, the potentials—and accordingly the correlation function— can be constrained only up to $k^*<92$ MeV/c (the upper limit of the range over which phase shifts were measured).
For higher values, extrapolations were applied. This method allows for a calculation of the wave function down to $r=0$, satisfying the necessary boundary conditions. This makes it possible to capture the full dynamics of the interaction including short distances.
\begin{figure}
\centering
\includegraphics[width=0.99\linewidth]{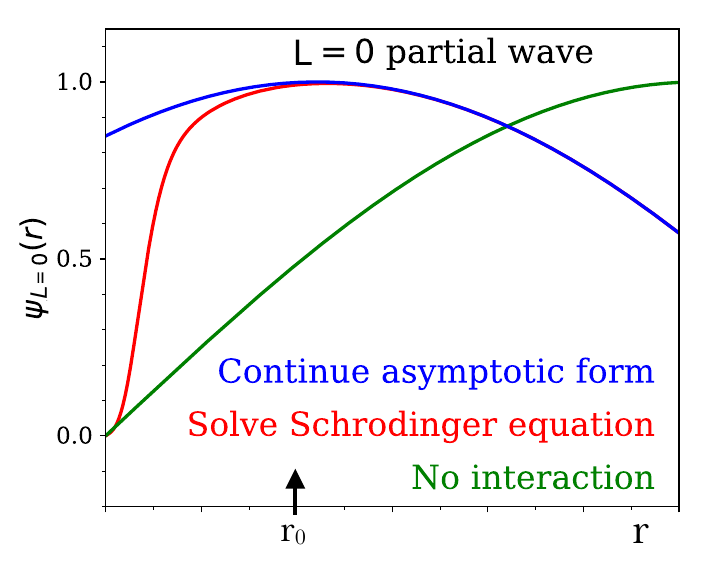}

\caption{Illustration of the construction of partial wave function dependant on range of interaction. In absence of potential (green) it behaves as  $\sin(k^*r)$. Adding the potential and solving the Schr\"odinger  equation results in a wave function with additional strength near the origin (red). As an approximation, one can use the asymptotic form, $\sin(k^*r+\delta)$ (blue). This depends only on the phase shift.} 
\label{fig:FunctionComparison}
\end{figure}

Figure \ref{fig:FunctionComparison} illustrates an example of a partial wave as a function of relative distance between particle and potential. As shown, the difference between the asymptotic form and the numerically solved wave function becomes significant at smaller values of $r$. In that region the asymptotic form extrapolate its solution whereas the exact calculations utilising the Schr\"odinger equation reduces the values of the function to zero at $r=0$. In the case of correlations involving the deuteron, it is essential to consider the size of the deuteron itself which increase the range of relative distance where the differences between two approaches became significant.

The correlation functions derived from Schr\"odinger based solutions of the wave function calculated in the CorAL package~\cite{coral, coralwww} are presented in the right panel of Fig.~\ref{fig:swave_2panel}. Unlike in the L-L approach, the functions show smooth behavior without any structure at very small $k^*$ even for small femtoscopic sources.

For the CorAL package, the potentials are chosen to be combinations of simple square wells. Combinations of spherical Coulomb waves are found that satisfy all the boundary conditions. The depths and widths of the wells were altered to best fit the experimental phase shifts. Potentials were found for total intrinsic spins of deuteron $S=0$ and $S=1$, and for orbital angular momenta, $L=0,1,2$. In reality, spin-orbit couplings provide splittings that depend on the total orbital angular momenta $J$. CorAL ignores this dependence, even though it is measured by experiment. For a given $S,L$ combination, CorAL averages phase shifts over the various values of $J$, with the averaging weighted by $2J+1$. Because the spin-orbit splittings between the phase shifts are rather small, one would expect this approximation to be quite reasonable.

Looking at Figure \ref{fig:FunctionComparison}, one might wonder if solving the Schr\"odinger equation for different potentials that produce the same phase shift would result in nearly identfical correlation functions. Since the phase shifts are the same and the boundary condition at $r=0$ is met, there might be limited flexibility in altering the wave function in between. Further, there is a sum rule involving the phase shifts. For any partial wave:
\begin{equation}
\int dr\left(|\psi(r)|^2-|\psi_0(r)|^2\right)=\frac{1}{2}\frac{d\delta}{dk^*},
\end{equation}
where $\psi_0$ is the wave function in the absence of the strong interaction. Thus, if two potential give both the same phase shifts and the same derivative of the phase shifts, the integral of $|\psi(r)|^2$ are identical within the range of $r_0$. This makes it even more difficult to distinguish between two potentials that both fit the same phase shifts. As a side note, it can be worth noting that in scattering theory phase shifts that differ by $\pi$ are indistinguishable. For example an attractive interaction with a bound state might result in a phase shift that starts at $\pi$ and falls with energy, whereas for a repulsive potential, the phase shift would typically start at zero and fall. The cases might have the same slope, $d\delta/dk^*$, and their phase shifts might differ by exactly $\pi$. However, the wave functions would differ significantly. For the attractive interaction, the wave function would have an additional node, this being necessary for the wave function to be orthogonal to the bound state.

\begin{figure}
\centering
\includegraphics[width=0.99\linewidth]{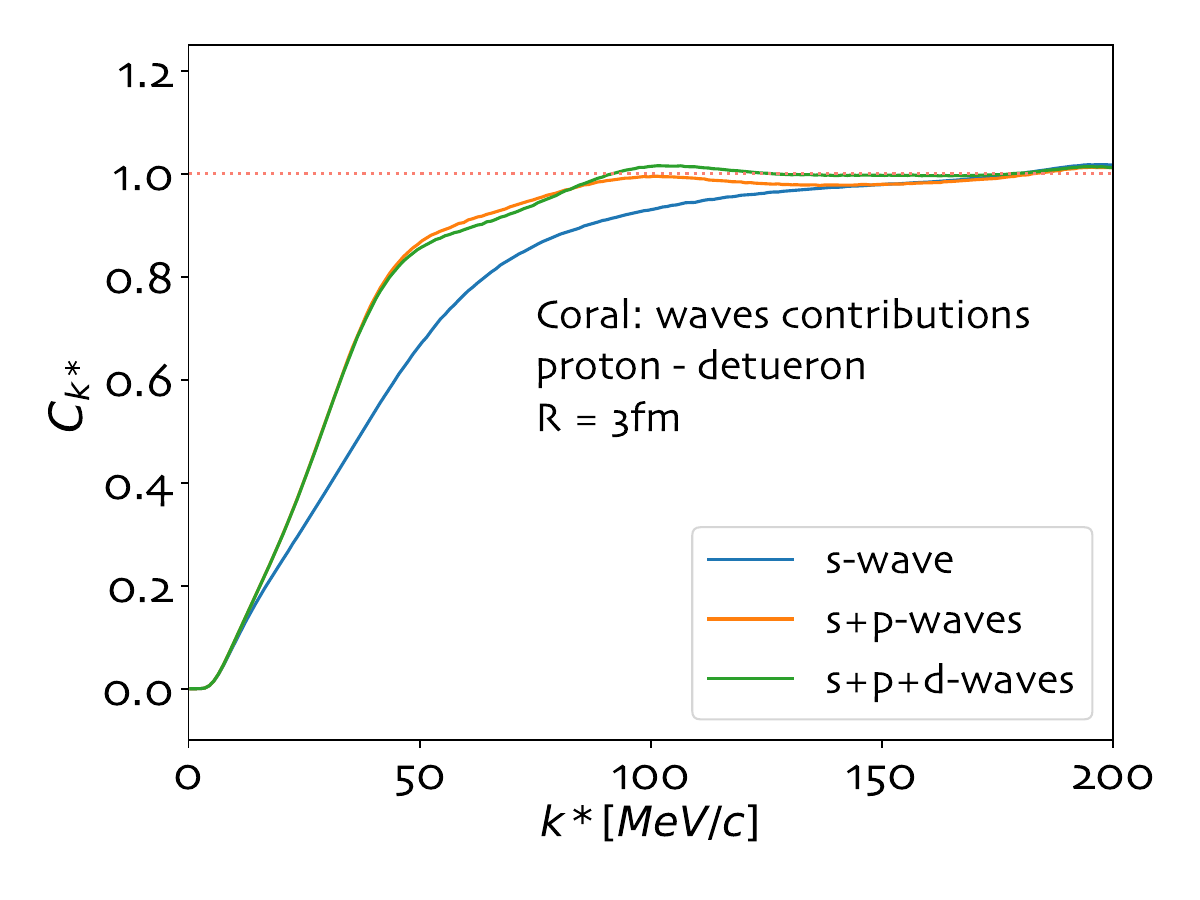}
\caption{Correlation functions of proton--deuteron pairs calculated with the CorAL package. The functions take into account Coulomb and strong interactions as well as partial wave contributions: s waves only (blue), s+p waves (orange) and s+p+d waves (green).}
\label{fig:WaveContribution}
\end{figure}

Another important aspect of describing strong interactions, particularly for \textit{p-d} systems, is the inclusion of not only s-wave but also 
the waves from higher orbital momentum $L$ =1 (p-waves) and $L$ =2 (d-waves). As demonstrated in~\cite{pd_theory_3body_Viviani2023, pd_ppcol_ALICE}, incorporating higher partial waves is crucial for accurately capturing the dynamics of \textit{p-d} interactions and understanding the properties of this system. An example of the impact of different partial waves in two-body approach is presented in Fig. \ref{fig:WaveContribution}. As shown in the figure, the contribution of the higher-order p-wave is significant and noticeably alters the nature of the interaction. Specifically, it reduces the strength of the interaction between the proton and deuteron, making it less repulsive compared to what is predicted by the s-wave alone, which as pointed in~\cite{pd_theory_3body_Viviani2023} is caused by the possible formation of the $^3$He bound state in the $S=1/2$ spin state. This highlights the importance of including higher partial waves, such as the p-wave, in accurately describing the dynamics of the \textit{p-d} system.

\begin{figure*}
\centering
\includegraphics[scale=0.38]{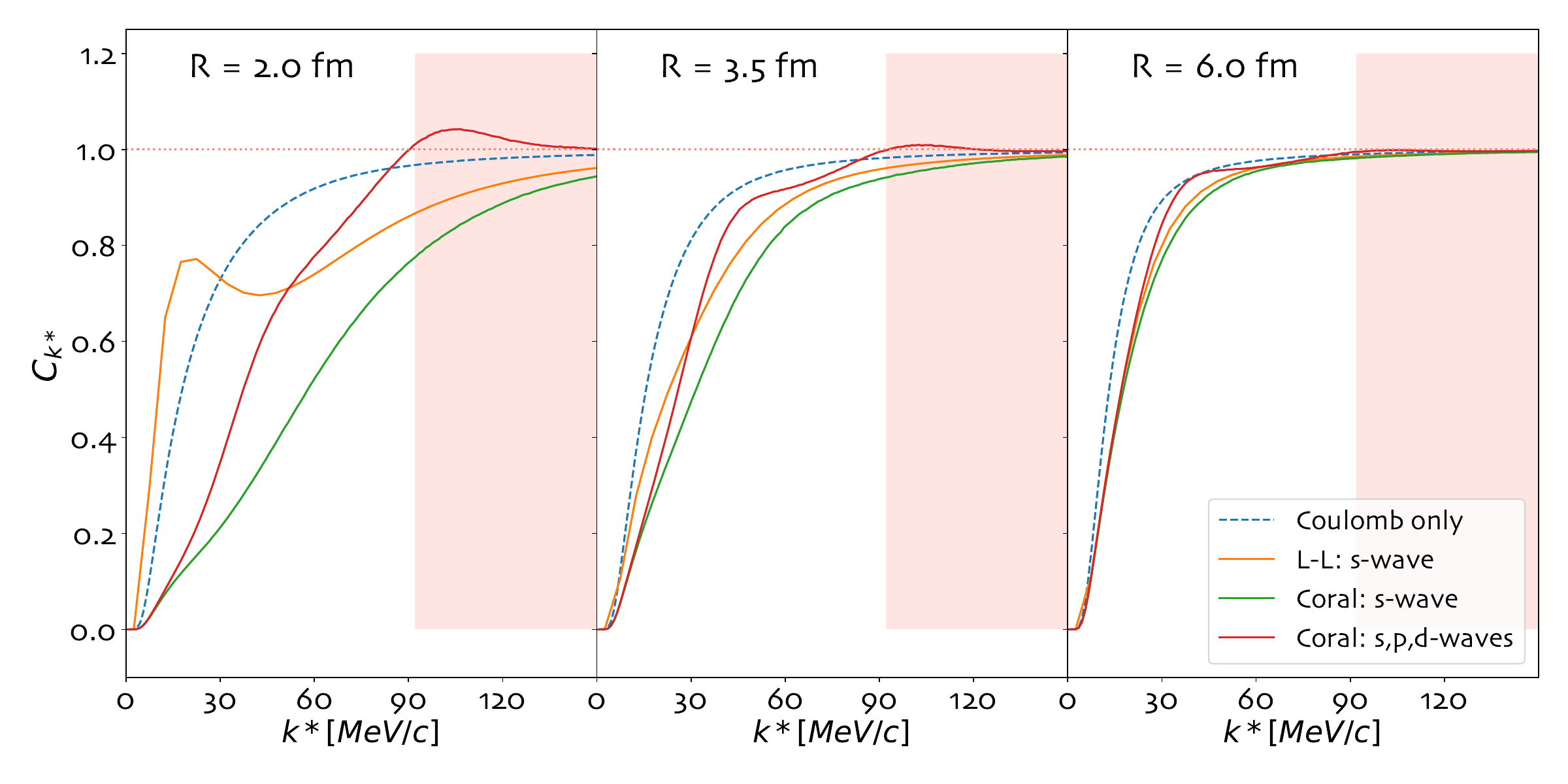}
\caption{The correlation function of proton--deuteron pairs calculated considering only the Coulomb interaction (dashed blue), Coulomb plus strong with s-wave contribution derived with the Lednick\'{y}-Lyuboshits formalism (denoted as L-L, solid orange line) and the CorAL package (solid green line), and s+p+d waves calculated with the CorAL package (solid red line). The functions are shown for different femtoscopic source sizes in three separate panels. The shaded area indicates the extrapolation solutions above the range of the exact phase shift measurements.}
\label{fig:comparison}
\end{figure*}

\section{Model comparison}
To effectively compare the two two-body methods for determining the \textit{p-d} theoretical correlation functions, specifically:

\begin{itemize} \item the asymptotic form in the Lednick\'{y}-Lyuboshits parametrization, using input scattering parameters from~\cite{Black1999} with the zero effective-range approximation, and 
\item the complete solution of the Schr\"odinger equation within the CorAL package~\cite{coral, coralwww}, based on the potential determined from phase shift measurements in~\cite{Black1999, improved_phaseshift}, \end{itemize}
the correlation functions are checked in thee different scales of femtoscopic Gaussian source and presented in Fig.~\ref{fig:comparison}. The dashed blue curve, represents the Coulomb only interaction, that serves as a baseline to highlight the impact of strong interactions on the shape of the functions. In the both methods, the s-waves of the strong interactions introduce a significant repulsive component to the correlation function. However, the discrepancies between the two models are clearly visible and especially pronounced in smaller femtoscopic source sizes (e.g., $R=2$ fm). In such functions where the dominant contribution come from particles at small relative distance the L-L model exhibits a characteristic peak that is absent in the calculations based on the exact solution of the Schr\"odinger equation. Since the key difference between the two-body approaches is the use of the asymptotic approximation, we attribute this peak and the difference of the magnitude of the s-waves contribution to that approximation. All the differences became smaller with growing system sizes.

In the Figs.~\ref{fig:WaveContribution} and ~\ref{fig:comparison} the Schr\"odinger based solutions show also the impact of the higher partial contributors, p- and d-waves. The two p-waves provide an especially visible attractive contribution to the correlation function. The characteristic "wiggles" visible in the CorAL calculations, which incorporate s-, p-, and d-waves, arise from the complexity of the potential and were also observed in~\cite{pd_ppcol_ALICE, pd_theory_3body_Viviani2023}. The impact of higher partial waves on the overall shape of the correlation function visibly decrease with growing femtoscopic source sizes.

\section{Conclusions}
In this paper, the model study of femtoscopic two-body \textit{p-d} correlation functions was presented. Two different approaches for determining the wave function were analyzed, the asymptotic approach of the L-L formalism and the exact solution of the Schr\"odinger equation. We made detailed comparisons between the two different approaches to describe the \textit{p-d} correlation functions, focusing on two-body systems where the expected source sizes are $\geq$ 2~fm. Our results indicate a significant impact of the asymptotic approximation on the description of the \textit{p-d} interactions, which can be pronounced due to the large size of the interacting pair and their strong effects of mutual interactions. The differences are obvious and not negligible especially for small femtoscopic sizes, and therefore the use of solutions with asymptotic approximations is more than questionable there. However, these differences decrease with increasing femtoscopic sizes of the system. Furthermore, we have investigated the contributions of different partial waves to the potential and the corresponding correlation functions, showing that the inclusion of higher-order waves - starting with the pronounced influence of the p-wave - is essential for a complete understanding of \textit{p-d} interactions at the femtoscopic scale.

We strongly recommend that, in any attempts to describe \textit{p-d} interactions especially in small femtoscopic sizes, a full solution of the Schr\"odinger equation should be employed. The differences between the two approaches investigated in this work should be captured withing the statistical significant experimental measurements (as seen, for example, in preliminary results from STAR \cite{STAR_pd} and HADES \cite{Stefaniak_pd}). We encourage further detailed studies on the impact of any approximations implemented in potential calculations, particularly in systems involving light-ions such as deuteron, helium and more composite objects.

\section{Acknowledgments}

We would like to express our gratitude to the entire "HBT camp" community for their insightful discussions and the inspiration that led to this paper. This work was supported by The Ohio State University President’s Postdoctoral Scholars Program, the U.S. Department of Energy, Office of Science, Office of Nuclear Physics, under Award Number DE-SC0020651, the Polish National Science Centre programme under agreement no. 2023/49/N/ST2/03525 and the IDUB YOUNG-PW programme no. 1820/96/Z01/2023.

\bibliographystyle{ieeetr}
\bibliography{main}

\begin{thebibliography}{10}

\bibitem{Oers1967}
W.~T. Van~Oers and K.~W. Brockman, Jr, ``{Phase-Shift Analysis of Elastic
  Nucleon--Deuteron Scattering.},''

\bibitem{Arvieux1974}
J.~Arvieux, ``{Phase-Shift Analysis of Elastic Proton-Deuteron Scattering Cross
  Sections and 3He Excited States},'' {\em Nuclear Physics A}, vol.~221, no.~2,
  pp.~253--268, 1974.

\bibitem{Huttel1983}
E.~Huttel, W.~Arnold, H.~Baumgart, H.~Berg, and G.~Clausnitzer, ``{Phase-Shift
  Analysis of pd Elastic Scattering Below Break-up Threshold},'' {\em Nuclear
  Physics A}, vol.~406, no.~3, pp.~443--455, 1983.

\bibitem{Kievsky1997}
A.~Kievsky, S.~Rosati, M.~Viviani, C.~Brune, H.~Karwowski, E.~Ludwig, and
  M.~Wood, ``{The Three-Nucleon System Near the N-d Threshold},'' {\em Physics
  Letters B}, vol.~406, no.~4, pp.~292--296, 1997.

\bibitem{Black1999}
T.~Black, H.~Karwowski, E.~Ludwig, A.~Kievsky, S.~Rosati, and M.~Viviani,
  ``Determination of proton-deuteron scattering lengths,'' {\em Physics Letters
  B}, vol.~471, no.~2, pp.~103--107, 1999.

\bibitem{femto2decades}
M.~A. Lisa, S.~Pratt, R.~Soltz, and U.~Wiedemann, ``Femtoscopy in relativistic
  heavy ion collisions: Two decades of progress,'' {\em Annual Review of
  Nuclear and Particle Science}, vol.~55, no.~1, pp.~357--402, 2005.

\bibitem{kp_in_pp}
S.~Acharya {\em et~al.}, ``Scattering studies with low-energy kaon-proton
  femtoscopy in proton-proton collisions at the lhc,'' {\em Phys. Rev. Lett.},
  vol.~124, p.~092301, Mar 2020.

\bibitem{kp_in_PbPb}
S.~Acharya {\em et~al.}, ``{Kaon–proton strong interaction at low relative
  momentum via femtoscopy in Pb–Pb collisions at the LHC},'' {\em Physics
  Letters B}, vol.~822, p.~136708, 2021.

\bibitem{barionantibarioninter}
S.~Acharya {\em et~al.}, ``Measurement of strange baryon–antibaryon
  interactions with femtoscopic correlations,'' {\em Physics Letters B},
  vol.~802, p.~135223, 2020.

\bibitem{STAR:2014shf}
L.~Adamczyk {\em et~al.}, ``{Beam-energy-dependent two-pion interferometry and
  the freeze-out eccentricity of pions measured in heavy ion collisions at the
  STAR detector},'' {\em Phys. Rev. C}, vol.~92, no.~1, p.~014904, 2015.

\bibitem{2023138145}
S.~Acharya {\em et~al.}, ``Accessing the strong interaction between $\lambda$
  baryons and charged kaons with the femtoscopy technique at the lhc,'' {\em
  Physics Letters B}, vol.~845, p.~138145, 2023.

\bibitem{Mrowczynski:2019yrr}
S.~Mr\'owczy\'nski and P.~S\l{}o\'n, ``{Hadron\textendash{}Deuteron
  Correlations and Production of Light Nuclei in Relativistic Heavy-ion
  Collisions},'' {\em Acta Phys. Polon. B}, vol.~51, no.~8, pp.~1739--1755,
  2020.

\bibitem{ant:2}
S.~Bazak, Sylwia ;~Mrówczyński, ``{Production of $ ^4Li$ and p-$^3He$
  correlation function in relativistic heavy-ion collisions},'' {\em The
  European Physical Journal A}, 2020.

\bibitem{ant:3}
S.~Mrówczyński and P.~Słoń, ``{Deuteron-Deuteron Correlation Function in
  Nucleus-Nucleus Collisions},'' {\em Physical Review C}, 2021.

\bibitem{ant:4}
S.~Mrówczyński and P.~Słoń, ``{Production of light nuclei at colliders -
  coalescence vs. thermal model},'' {\em European Physical Journal Special
  Topics}, 2020.

\bibitem{prod1}
E.~V. Shuryak, ``Quantum chromodynamics and the theory of superdense matter,''
  {\em Physics Reports}, vol.~61, no.~2, pp.~71--158, 1980.

\bibitem{prod2}
P.~Braun-Munzinger and J.~Stachel, ``{Production of strange clusters and
  strange matter in nucleus-nucleus collisions at the AGS},'' {\em J. Phys. G},
  vol.~21, pp.~L17--L20, 1995.

\bibitem{prod3}
P.~Braun-Munzinger and J.~Stachel, ``{Particle ratios, equilibration, and the
  QCD phase boundary},'' {\em J. Phys. G}, vol.~28, pp.~1971--1976, 2002.

\bibitem{prod4}
A.~Baltz, C.~Dover, S.~Kahana, Y.~Pang, T.~Schlagel, and E.~Schnedermann,
  ``{Strange cluster formation in relativistic heavy ion collisions},'' {\em
  Phys. Lett. B}, vol.~325, 1994.

\bibitem{SHMmodel}
V.~Vovchenko and H.~Stoecker, ``{Thermal-FIST: A package for heavy-ion
  collisions and hadronic equation of state},'' {\em Computer Physics
  Communications}, vol.~244, pp.~295--310, 2019.

\bibitem{CoalInHIC}
R.~Scheibl and U.~Heinz, ``{Coalescence and flow in ultrarelativistic heavy ion
  collisions},'' {\em Phys. Rev. C}, vol.~59, pp.~1585--1602, Mar 1999.

\bibitem{hbt1954}
R.~H. Brown and R.~Twiss, ``Lxxiv. a new type of interferometer for use in
  radio astronomy,'' {\em The London, Edinburgh, and Dublin Philosophical
  Magazine and Journal of Science}, vol.~45, no.~366, pp.~663--682, 1954.

\bibitem{Goldhaber1960}
G.~Goldhaber, S.~Goldhaber, W.~Lee, and A.~Pais, ``Influence of bose-einstein
  statistics on the antiproton-proton annihilation process,'' {\em Phys. Rev.},
  vol.~120, pp.~300--312, Oct 1960.

\bibitem{KOPYLOV1972}
V.~Grishin, G.~Kopylov, and M.~Podgoretsky, ``Correlations of identical
  particles emitted by highly excited nuclei,'' {\em Sov. J. Nucl. Phys.},
  1972.

\bibitem{KOPYLOV1974}
G.~Kopylov, ``Like particle correlations as a tool to study the multiple
  production mechanism,'' {\em Physics Letters B}, vol.~50, no.~4,
  pp.~472--474, 1974.

\bibitem{sinyukov1}
Y.~M. Sinyukov, ``Spectra and correlations in locally equilibrium hadron and
  quark-gluon systems,'' {\em Nucl. Phys. A}, vol.~566, p.~589, 1994.

\bibitem{Koonin:1977fh}
S.~E. Koonin, ``Proton pictures of high-energy nuclear collisions,'' {\em Phys.
  Lett. B}, vol.~70, pp.~43--47, 1977.

\bibitem{pratt:1986}
S.~Pratt, ``Pion interferometry of quark-gluon plasma,'' {\em Phys. Rev. D},
  vol.~33, pp.~1314--1327, Mar 1986.

\bibitem{Broniowski:2008vp}
W.~Broniowski, M.~Chojnacki, W.~Florkowski, and A.~Kisiel, ``{Uniform
  Description of Soft Observables in Heavy-Ion Collisions at $\sqrt{s_{\rm NN}}
  = 200$~GeV},'' {\em Phys. Rev. Lett.}, vol.~101, p.~022301, 2008.

\bibitem{Acharya:2017qtq}
S.~Acharya {\em et~al.}, ``{Kaon femtoscopy in Pb-Pb collisions at
  $\sqrt{s_{\rm{NN}}}$ = 2.76 TeV},'' {\em Phys. Rev. C}, vol.~96, no.~6,
  p.~064613, 2017.

\bibitem{Acharya:2020nyr}
S.~Acharya {\em et~al.}, ``{Pion-kaon femtoscopy and the lifetime of the
  hadronic phase in Pb$-$Pb collisions at $\sqrt{s_{\rm{NN}}}$ = 2.76 TeV},''
  {\em Phys. Lett. B}, vol.~813, p.~136030, 2021.

\bibitem{pd_theory_3body_Viviani2023}
M.~Viviani, S.~K\"onig, A.~Kievsky, L.~E. Marcucci, B.~Singh, and O.~V. Doce,
  ``Role of three-body dynamics in nucleon-deuteron correlation functions,''
  {\em Phys. Rev. C}, vol.~108, p.~064002, Dec 2023.

\bibitem{pd_ppcol_ALICE}
S.~Acharya {\em et~al.}, ``{Exploring the strong interaction of three-body
  systems at the LHC},'' {\em Phys. Rev. X}, 2023.

\bibitem{Lednicky1981}
R.~Lednicky and V.~L. Lyuboshits, ``{Final State Interaction Effect on Pairing
  Correlations Between Particles with Small Relative Momenta},'' {\em Yad.
  Fiz.}, vol.~35, pp.~1316--1330, 1981.

\bibitem{nonidentical_femtoscopy_Lednicky:2009zza}
R.~Lednicky, ``{Femtoscopic correlations of nonidentical particles},'' {\em
  Acta Phys. Polon. B}, vol.~40, pp.~1145--1154, 2009.

\bibitem{clebish_gordon}
D.~Mihaylov, V.~{Mantovani Sarti}, O.~Arnold, L.~Fabbietti, B.~Hohlweger, and
  A.~Mathis, ``A femtoscopic correlation analysis tool using the
  schr{\"o}dinger equation (cats),'' {\em European Physical Journal C},
  vol.~78, May 2018.
\newblock Publisher Copyright: {\textcopyright} 2018, The Author(s).

\bibitem{Stefaniak_pd}
M.~Stefaniak, ``{Proton-cluster femtoscopy with the HADES experiment},'' {\em
  EPJ Web Conf.}, vol.~296, p.~02001, 2024.

\bibitem{improved_phaseshift}
W.~Tornow, A.~Kievsky, and H.~Witała, ``{Improved Proton-Deuteron Phase-Shift
  Analysis Above the Deuteron Breakup Threshold and the Three-Nucleon
  Analyzing-Power Puzzle},'' {\em Few-Body Systems}, vol.~32, no.~9, 2002.

\bibitem{coral}
D.~A. Brown, A.~Enokizono, M.~Heffner, R.~Soltz, P.~Danielewicz, and S.~Pratt,
  ``{Imaging three dimensional two-particle correlations for heavy-ion reaction
  studies},'' {\em Phys. Rev. C}, vol.~72, p.~054902, 2005.

\bibitem{coralwww}
S.~Pratt, ``{Github: Coral project
  https://github.com/scottedwardpratt/coral},'' 2024.

\bibitem{STAR_pd}
K.~Mi, ``{Femtoscopy of Proton, Light Nuclei, and Strange Hadrons in Au$+$Au
  Collisions at STAR},'' {\em Acta Phys. Polon. Supp.}, vol.~16, no.~1,
  pp.~1--A91, 2023.

\end{thebibliography}

\end{document}